\documentclass[aps,prl,superscriptaddress,onecolumn,10pt]{revtex4-1}
\usepackage{graphicx,color}
\usepackage{array,url,kantlipsum}
\usepackage{bm}
\usepackage{amssymb,amsmath,amsfonts, mathrsfs}

\newcommand{\be}{\begin{equation}}
\newcommand{\ee}{\end{equation}}
\newcommand{\ba}{\begin{array}}
\newcommand{\ea}{\end{array}}
\newcommand{\bqa}{\begin{eqnarray}}
\newcommand{\eqa}{\end{eqnarray}}

\begin{document}
\begin{flushright}
USTC-ICTS-19-08\\
\end{flushright}


\title{Decays of $X(3872)$ to $\chi_{cJ}\pi^0$ and $J/\psi\pi^+\pi^-$}



\author{Zhi-Yong Zhou}
\email[]{zhouzhy@seu.edu.cn}
\affiliation{School of Physics, Southeast University, Nanjing 211189,
P.~R.~China}

\author{Meng-Ting Yu}
\affiliation{School of Physics, Southeast University, Nanjing 211189,
P.~R.~China}

\author{Zhiguang Xiao}
\email[Corresponding author:]{xiaozg@ustc.edu.cn}
\affiliation{Interdisciplinary Center for Theoretical Study, University of Science
and Technology of China, Hefei, Anhui 230026, China}


\date{\today}

\begin{abstract}
By describing the $X(3872)$ using the extended Friedrichs scheme, in
which $D\bar D^*$ is considered as the dominant component,  we
calculate the decay rates of the $X(3872)$ to $\pi^0$ and a $P$-wave
charmonium $\chi_{cJ}$ state with $J=0,1$, or $2$, and the rate of its
decay to $J/\psi\pi^+\pi^-$ with the help of Barnes-Swanson model,
where $\pi^+\pi^-$ are assumed to be produced via an intermediate
$\rho$ state. This calculation shows that the decay rate of $X(3872)$
to $\chi_{c1}\pi^0$ is one order of magnitude smaller than its decay
rate to $J/\psi\pi^+\pi^-$ and the decay widths of
$X(3872)\to\chi_{cJ}\pi^0$ for $J=0,1,2$ are of the same order.

\end{abstract}


\maketitle
Discovery of the narrow hadron state $X(3872)$, first observed by the
Belle Collaboration in 2003~\cite{Choi:2003ue} and soon confirmed by
the CDF, \textsl{BABAR}, and D0
Collaborations~\cite{Acosta:2003zx,Aubert:2004fc,Abazov:2004kp},
challenges the prediction of quark model and arouses enormous
experimental explorations and theoretical studies, as reviewed by Refs.~\cite{Guo:2017jvc,Chen:2016qju,Lebed:2016hpi}.
Recently, the BESIII Collaboration searched for the $X(3872)$ signals in
$e^+e^-\to\gamma\chi_{cJ}\pi^0$ ($J=0,1,2$) and reported an
observation of $X(3872)\to\chi_{c1}\pi^0$ with a ratio of
branching fractions~\cite{Ablikim:2019soz}
\bqa
\frac{\mathcal{B}(X(3872)\to\chi_{c1}\pi^0)}{\mathcal{B}(X(3872)\to J/\psi\pi^+\pi^-)}=0.88^{+0.33}_{-0.27}\pm 0.10.
\eqa
They also set $90\%$ confidence level upper limits on the
corresponding ratios for the decays
to $\chi_{c0}\pi^0$ and $\chi_{c2}\pi^0$ as 19 and 1.1, respectively.
Soon after, the Belle Collaboration made a search for $X(3872)$ in $B^+\to \chi_{c1}\pi^0K^+$ but did not find a significant signal of $X(3872)\to\chi_{c1}\pi^0$. They reported an upper limit\cite{Bhardwaj:2019}
\bqa
\frac{\mathcal{B}(X(3872)\to\chi_{c1}\pi^0)}{\mathcal{B}(X(3872)\to J/\psi\pi^+\pi^-)}<0.97
\eqa
at $90\%$ confidence level.

The ratio of $X(3872)$ decaying to $\chi_{cJ}\pi^0$ with $J=0,1,2$ is
suggested to be sensitive to the internal structure of $X(3872)$ in
Ref.~\cite{Dubynskiy:2007tj}, and the ratios of decay rates are
estimated to be $\Gamma_0:\Gamma_1:\Gamma_2=0:2.7:1$ when assuming the
$X(3872)$ as a traditional charmonium state or
$\Gamma_0:\Gamma_1:\Gamma_2=2.88:0.97:1$ as a four-quark state.
Several other calculations in a similar spirit are also carried
out in
Refs.~\cite{Fleming:2008yn,Fleming:2011xa,Mehen:2015efa,Guo:2010ak,Dong:2009yp}
based on the effective field theory~(EFT) approach. Another popular
picture of $X(3872)$ is that it is a dynamically generated state by
the strong interaction between the $\chi_{c1}(2P)$ $c\bar c$ bare state and the continuum states
such as $D\bar D^*$,  which have OZI-allowed coupling to  $c\bar
c$~\cite{Coito:2012vf,Takizawa:2012hy,Takeuchi:2014rsa,Zhou:2017dwj}.
As a result, considering only the formation of $X(3872)$, the
wave function of $X(3872)$ at this point mainly contains $c\bar c$ and
those OZI-allowed components, in which $D\bar D^*$ were found to be
dominant.  This picture may overcome the problem of prompt
production~\cite{Bignamini:2009sk} and radiative
decay~\cite{Swanson:2004pp,Dong:2009uf} met by the pure molecule
explanation. Since the couplings of the $\chi_{cJ} \pi^0$ to $c\bar c$
component are too small and  can be ignored while their coupling to
$D\bar D^*$ components are OZI-allowed, it is expected that the decays
of $X(3872)$ to $\chi_{cJ} \pi^0$  are contributed mainly through the
dominant components $D\bar D^*$. This point of view was also adopted
in \cite{Swanson:2006st} in discussing the $X(3872)\to J/\psi\pi\pi$
decay. Thus, a calculation of the decay from
this point of view is in demand. This picture is different from the
effective field theory approach~\cite{Dong:2009yp} from a pure
molecule point of view, where $D\bar D^*$, $J/\psi\rho$, $J/\psi
\omega$ are treated on the same footing in the wave function of
$X(3872)$ while the $\chi_{c1}(2P)$ $c\bar c$ component is not considered.

In this paper we would undertake a new calculation just in the above picture from the
constituent quark point of view and consider the $D\bar D^*$ as the
main contribution to the decay. In principle, calculations at the
constituent quark level have proved to be successful in understanding
the mass spectrum of most meson states and the model parameters have
been determined to high accuracy, such as in Godfrey-Isgur~(GI)
model~\cite{Godfrey:1985xj}. Furthermore, the constituent quark models
use the wave functions of the meson states to represent the dynamical
structure of the state rather than regard them as a pointlike state,
which also naturally suppress the divergences in the large momentum
region.

The theoretical basis of this work is that $X(3872)$ state
automatically emerges in the extended Friedrichs scheme and can
be expressed as the combination of the $c\bar c$ components and the continuum
components such as $D\bar D^*$, in which the $D\bar D^*$ component
is dominant~\cite{Zhou:2017dwj}.
This picture has proved successful in obtaining the mass and width and
the isospin-breaking effects of the $X(3872)$
decays~\cite{Zhou:2017txt}, and another calculation with the similar
spirit also indicates the reasonability of this scheme~\cite{Giacosa:2019zxw}. This approach can be extended to
discuss the decays to $\chi_{cJ}\pi^0$ processes by considering one of the  final states being a $P$-wave state. Since the dominant
continuum components is $D\bar D^*$, and the pure $c\bar c$ contribution
is OZI suppressed, we consider only the contribution from $D\bar D^*$ component of
$X(3872)$ to the decay. Since the $D\bar D^*$ component could be separated
into $S$-wave and $D$-wave parts, we need to calculate the amplitude
of these different angular momentum components
 to the $P$-wave final $\chi_{cJ}\pi^0$. This can be achieved by
the Barnes-Swanson
model~\cite{Barnes:1991em,Barnes:1999hs,Barnes:2000hu,Wong:2001td}. This model has
been used in studying the heavy meson scattering
~\cite{Wang:2018pwi,Liu:2014eka}. With these partial-wave amplitude,
the decay rates of $X(3872)$ to $\chi_{cJ}\pi^0$, $J/\psi\rho$, and
$J/\psi\omega$ could be calculated by combining the previous result
from the Friedrichs model scheme, and thus the branching fractions could be
obtained.  In this calculation, there is no free parameter introduced
since all the parameters are the input of the GI model or have been
determined by obtaining the correct $X(3872)$
pole~\cite{Zhou:2017dwj}. However, since this
calculation has some model dependence, we  would not expect this
approach to give a precise result of the decay width, but just an order of
magnitude estimate.  Nevertheless, we found that in this calculation the
decay rates of $X(3872)$ to the $\chi_{cJ}\pi^0$ are one order of
magnitude smaller than its decays to $J/\psi\pi^+\pi^-$.

The calculation is based on our previous result where, in the extend Friedrichs scheme~\cite{Xiao:2016mon,Xiao:2016wbs}, the
$X(3872)$ state is dynamically generated by the coupling between the
bare discrete $\chi_{c1}(2P)$ state and the continuum $D\bar D^*$ and $D^*\bar D^*$ states~\cite{Zhou:2017dwj}, and its wave function  could be explicitly written down as
\begin{align}
&|X\rangle=N_B\Big(|c\bar c\rangle+\int_{M_{00}}^\infty \mathrm{d}E
\sum_{l,s}\frac{f^{00}_{ls}(E)} {z_X-E}(|E\rangle_{ls}^{D^0\bar
D^{0*}}+C.C.
)\nonumber\\&+\int_{M_{+-}}^\infty
\mathrm{d}E\sum_{l,s}\frac{f^{+-}_{ls}(E)}
{z_X-E}(|E\rangle_{ls}^{D^+D^{-*}}+C.C.)+\cdots\Big) ,
\label{eq:X3872-wave-funtion}
\end{align}
where $C.C.$ means the
corresponding charge conjugate state, $|c\bar c\rangle$ denotes the bare $\chi_{c1}(2P)$ state and
$|E\rangle_{ls}^n=\sqrt{\mu k}|k, j\sigma,ls\rangle$ denotes  the
two-particle $``n"$ state~(``$n$" denotes the species of the
continuum state) with the reduced mass $\mu$, the
magnitude of one-particle three-momentum $k$ in their c.m. frame,
total spin $s$, relative orbital angular momentum $l$, total angular
momentum $j$, and its third component $\sigma$.
The coupling form factors $f^{00}_{ls}$ and $f^{+-}_{ls}$ could also
be written down explicitly by using the quark pair creation
model~\cite{Micu:1968mk,Blundell:1995ev} and the wave functions from
the quark potential models, such as the GI
model~\cite{Godfrey:1985xj}.  $M_{00}$ and $M_{+-}$ in the integral
limits are the threshold energies of $D^0\bar D^{0*}$ and $D^+\bar
D^{-*}$ respectively. $z_X$ is the dynamically generated $X(3872)$ pole position, one of the zero
points $\eta(z)$, the inverse of the resolvent,  and $N_B=\eta'(z_X)^{-1/2}$ is the
normalization factor, where $\eta(z)$ is defined as
$\eta(z)=z-E_0-\sum_{n,l,s}\int_{E_{n,th}}^\infty \mathrm{d}E \frac{|f^n_{ls}(E)|^2}{z-E}.
$
The $\cdots$ represents other continuous states such as $D^*\bar D^*$,
but the compositeness of $D^*\bar D^*$ continua is about 0.4 percent
such that their contribution to this calculation is tiny and could be omitted.

In general, the transition rate for a
single-particle state $\alpha$ decaying into a two-particle state $\beta$~(including  particle $\beta_1$ and particle $\beta_2$) could be represented as
$d\Gamma(\alpha\to \beta)=2\pi|M_{\beta\alpha}|^2\delta^4(p_{\beta_1}+p_{\beta_2}-p_\alpha)d^3\vec p_{\beta_1}d^3\vec p_{\beta_2}$ where $M_{\beta\alpha}$ is the transition amplitude.
In a nonrelativistic approximation, the partial decay width can be
represented as
\bqa
\Gamma(\alpha\to\beta)=\sum_{l's'}2\pi |M_{l's'}|^2\mu' k'=\sum_{l's'}2\pi |F_{l's'}|^2
\eqa where $M_{l's'}$ is the partial-wave decay amplitude, $\mu'$ is
the reduced mass of two-particle state $\beta$,  $k'$ is the magnitude of three-momentum of one particle in their c.m. frame, and $F_{l's'}$ is the decay amplitude with the phase space factor $\sqrt{\mu' k'}$ absorbed in.

To calculate the hadronic decays of the $X(3872)$, e.g. to $\chi_{cJ}\pi^0$ for $J=0, 1, 2$, the partial-wave amplitude reads
\begin{align}
&F_{l's'}={_{l's'}}\langle \chi_{cJ}\pi^0|H_I|X(3872)\rangle=N_B\Big({^{\chi_{cJ}\pi^0}_{\ ~ l's'}}\langle E'|H_I|c\bar c\rangle\nonumber\\
&+\int_{M_{00}}^\infty \mathrm{d}E
\sum_{l,s}\frac{f^{00}_{ls}(E)} {z_X-E}({^{\chi_{cJ}\pi^0}_{\ ~ l's'}}\langle E'|H_I|E\rangle_{ls}^{D^0\bar
D^{0*}}+C.C.
)\nonumber
\\ &+\int_{M_{+-}}^\infty
\mathrm{d}E\sum_{l,s}\frac{f^{+-}_{ls}(E)}
{z_X-E}({^{\chi_{cJ}\pi^0}_{\ ~ l's'}}\langle
E'|H_I|E\rangle_{ls}^{D^+D^{-*}}+C.C.)\nonumber\\
&+\cdots\Big)
\label{eq:decay-amplitude}
\end{align}
where $C.C.$ means the matrix element from the corresponding charge conjugate
state. Once the matrix elements for $D\bar D^*\to \chi_{cJ}\pi^0$ with total
angular momentum $j=1$ are obtained, the partial decay widths and
branching ratios could be obtained directly.
In general, the hadron-hadron interaction matrix element of $AB\to CD$ is expressed as
\begin{align}
&{^{\ n'}_{ l's'}}\langle E'|H_I|E\rangle_{ls}^{n}=\delta{(E'-E)}\mathcal{M}^j_{l's'n',lsn}
\end{align}
and the partial-wave amplitude reads
\begin{align}
&\mathcal{M}^j_{l's'n',lsn}=\sqrt{\mu k\mu'k'}\sum_{\nu\nu'mm'\sigma_A\sigma_B\sigma_C\sigma_D}\nonumber\\
&\times
\langle j_A\sigma_A j_B\sigma_B|s\nu\rangle \langle s\nu lm|j\sigma\rangle  \langle j_C\sigma_C j_D\sigma_D|s'\nu'\rangle \langle s'\nu' l'm'|j\sigma\rangle\nonumber\\
&\times\int d\Omega_k\int d\Omega_{k'}\mathcal{M}_{\vec k'\sigma_C,-\vec k'\sigma_D;\vec k\sigma_A,-\vec k\sigma_B}
Y_l^m(\hat k)Y_{l'}^{m'*}(\hat k')
\end{align}
where $\nu$ is the third component of total spin $s$. The symbols with
primes represent the ones for the final states.

A simple model for calculating the scattering amplitude
$\mathcal{M}_{\vec k'\sigma_C,-\vec k'\sigma_D;\vec k\sigma_A,-\vec
k\sigma_B}$ is the Barnes-Swanson
model~\cite{Barnes:1991em,Barnes:1999hs,Barnes:2000hu,Wong:2001td}, which
evaluates the lowest~(Born) order $T$-matrix element between two-meson
scattering states by considering the interaction between the quarks or
antiquarks inside the scattering mesons.
In the $q_a (\bar q_a) +q_b (\bar q_b)\to q_{a'} (\bar q_{a'})+ q_{b'} (\bar
q_{b'})$ quark(antiquark) transitions, the initial and final momenta
are denoted as $\vec a\vec b\to \vec a'\vec b'$. It is convenient to define $\vec q=\vec a'-\vec a$, $\vec p_1=(\vec a'+\vec a)/2$, $\vec p_2=(\vec b'+\vec b)/2$.

In general, six kinds of interactions, the spin spin, color Coulomb,
linear, one gluon exchange~(OGE), spin orbit, linear spin orbit, and
tensor interactions, are considered, which is similar to the
interaction potential terms in obtaining the mass spectrum and the
meson wave functions in the GI model. Thus, they are consistent with
the calculations of the extended Friedrichs scheme to determine the wave function of the
$X(3872)$.

Four kinds of diagrams are considered, among which the quark-antiquark
interactions are denoted as $Capture_1$, $Capture_2$, and the
quark-quark(antiquark-antiquark) interactions are denoted as
$Transfer_1$, and $Transfer_2$. To reduce the so-called
``prior-post" ambiguity, the four ``post'' diagrams are considered
similarly and averaged to obtain the final result. For more details on the
calculation of the model, the readers are referred to the original
papers~\cite{Barnes:1991em,Barnes:2000hu,Wong:2001td}.

By standard derivation, one could obtain the partial-wave scattering
amplitude for each diagram with only meson $C$ being a
$P$-wave state using
\begin{align}
&\mathcal{M}^1_{l'j_C,lj_B}=\sqrt{\mu k\mu' k'}\sum_{mm'm_{l_C}}
\langle j_B-m lm|10\rangle \nonumber\\
& \times\langle j_C -m' l'm'|10\rangle \langle l_C m_{l_C} s_C (-m'-m_{l_C})|j_C -m'\rangle \nonumber\\
&\times  \langle \phi_{14}\phi_{32}|\phi_{12}\phi_{34}\rangle\langle \omega_{14}\omega_{32}|H_C|\omega_{12}\omega_{34}\rangle\nonumber\\
&\times\int d\Omega_k\int d\Omega_{k'}\langle \chi_C\chi_D|I_{Space}^{m_{l_C}}[\vec k,\vec k']|\chi_A\chi_B\rangle Y_l^{m}(\hat k)Y_{l'}^{m'*}(\hat k')
\end{align}
where  $\langle \phi_{14}\phi_{32}|\phi_{12}\phi_{34}\rangle$ is the
flavor factor, and $\langle
\omega_{14}\omega_{32}|H_C|\omega_{12}\omega_{34}\rangle$ the color
factor, which is $-4/9$ and $4/9$ for interactions of $q\bar q$ and $q
q$ respectively. $\chi_A$ represents the spin wave function of meson $A$. The space integral
\begin{align}\label{spacefactor}
&I_{Space}^{m_{l_C}}[\vec k,\vec k']=\int d^3p\int d^3q  \psi^A_{000}(\vec{p}_A)\psi^B_{000}(\vec{p}_B)\nonumber\\ &\psi^{C*}_{01m_{l_C}}(\vec{p}_C)\psi^{D*}_{000}(\vec{p}_D)T_{fi}(\vec q,\vec p_1,\vec p_2)
\end{align}
where $\psi_{n_rLm_L}(\vec p_r)$ is the wave
function for the bare meson state, with $n_r$ being the radial quantum
number, $L$  the relative angular momentum of the quark and
antiquark, $m_L$ its third component, and $\vec p_r$ is the relative momentum of quark and antiquark in the meson.
The quark interactions involved in this calculation are
\bqa
T_{fi}(\vec q,\vec p_1,\vec p_2)=\left\{\begin{array}{cc}
                -\frac{8\pi\alpha_s}{3m_1m_2}[\vec S_1\cdot\vec S_2] & \mathrm{Spin-spin}\\
                \frac{4\pi\alpha_s}{q^2}\mathbf{\textit{I}} & \mathrm{Coulomb} \\
                \frac{6\pi b}{q^4}\mathbf{\textit{I}} & \mathrm{Linear}\\
                \frac{4i\pi\alpha_s}{q^2}\{\vec S_1\cdot[\vec q\times(\frac{\vec p_1}{2m_1^2}-\frac{\vec p_2}{m_1m_2})]+\vec S_2\cdot[\vec q\times(\frac{\vec p_1}{m_1m_2}-\frac{\vec p_2}{2m_2^2})]\} & \mathrm{OGE\ spin-orbit}\\
                -\frac{3i\pi b}{q^4}[\frac{1}{m_1^2}\vec S_1\cdot(\vec q\times\vec p_1)-\frac{1}{m_2^2}\vec S_2\cdot(\vec q\times\vec p_2)] & \mathrm{Linear\ spin-orbit}\\
                \frac{4\pi\alpha_s}{m_1m_2q^2}[\vec S_1\cdot\vec q\vec S_2\cdot\vec q-\frac{1}{3}q^2\vec S_1\cdot\vec S_2] & \mathrm{OGE\  tensor}
              \end{array}\right.
\eqa
where $\alpha_s=\sum_k \alpha_k e^{-\gamma_k  q^2}$ as the parametrization form in the GI model. $m_1$ and $m_2$ are the masses of the two interacting quarks.

Similarly, one could obtain the decay amplitude of $X(3872)\to
J/\psi\rho$ and $J/\psi\omega$, which is simpler because there are only
$S$-wave states involved in the scattering amplitudes $\mathcal{M}_{l's'n',lsn}$.

As we analyze the properties of $X(3872)$, we use the famous GI model
as input. The wave functions of all the bare meson states have been
determined in the GI model. Furthermore, the Barnes-Swanson model does
not adopt any new parameter since the quark-quark interaction terms
share the same form as the GI model. The whole calculation has only
one free parameter, the quark pair creation strength $\gamma$, which
is determined by requiring $z_{X(3872)}=3.8716$GeV. The running
coupling constant is chosen as
$\alpha_s(q^2)=0.25e^{-q^2}+0.15e^{-\frac{q^2}{10}}+0.20e^{-\frac{q^2}{1000}}$,
and the quark masses are $m_u=0.2175$GeV, $m_d=0.2225$GeV,
$m_c=1.628$GeV, $b=0.18$, and $\gamma\simeq 4.0$. There is a technical
difficulty in the numerical calculation. To obtain the partial-wave
scattering amplitude, one encounters a ten-dimensional integration,
six for the momentum variables and four for the partial-wave
decomposition, which is not able to be calculated accurately by the
programme. To get around this difficulty, we make an approximation by
using the simple harmonic oscillator wave function to represent
the four involved mesons with their effective radii equal to the
rms radii calculated from the wave functions of GI model. In such a
simplification, the space overlap function of Eq.~(\ref{spacefactor})
could be integrated out analytically~\cite{Barnes:1999hs,Wong:2001td}.
Then, the
partial-wave integration is only four dimensional and can be evaluated numerically.

\begin{figure}[t]%
\begin{center}%
\includegraphics[height=40mm]{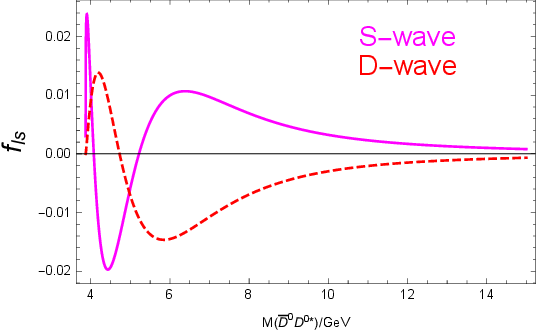}
\caption{\label{fig:X3872SDwave} $S$-wave term~(solid) and $D$-wave one~(dashed) of coupling form factors for $D^0\bar D^{0*}$ components.}
\end{center}%
\end{figure}

\begin{figure}[t]%
\begin{center}%
\includegraphics[height=40mm]{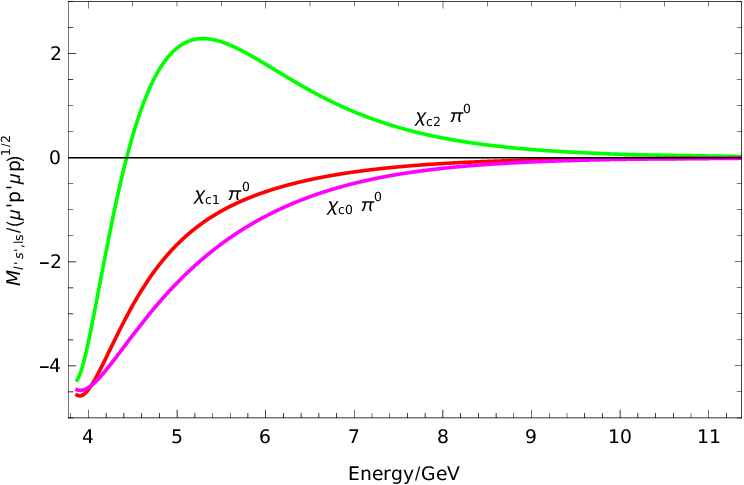}
\hspace{1cm}\includegraphics[height=40mm]{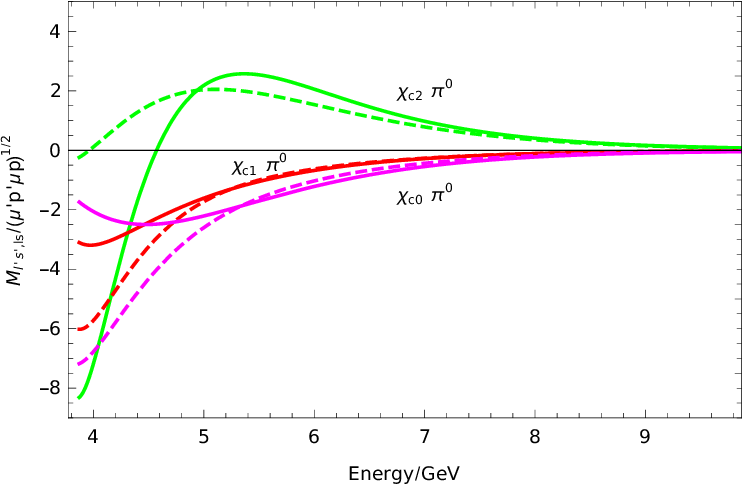}

\caption{\label{fig:chicJ} The scattering amplitudes without the phase
space factors of $D^0\bar D^{0*}\to \chi_{c0}\pi^0$,
$\chi_{c1}\pi^0$, $\chi_{c2}\pi^0$. The right one
shows the prior (solid) and the post (dashed) contributions to the
amplitudes. The left one shows
the averaged amplitudes.}
\end{center}%
\end{figure}

The wave function of $X(3872)$ has the $S$-wave and $D$-wave $D\bar
D^*$ components as shown in Fig.~\ref{fig:X3872SDwave}, both of which
could, in principle, transit to the final $P$-wave $\chi_{cJ}\pi^0$
state. However, the $S$-wave components contribute dominantly, and
their partial-wave scattering amplitudes to $P$-wave $\chi_{cJ}\pi^0$
states  are shown in Fig.~\ref{fig:chicJ}.

\begin{figure}[t]%
\begin{center}%
\includegraphics[height=40mm]{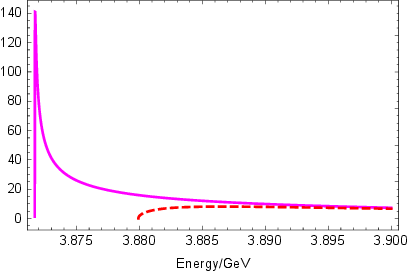}
\caption{\label{fig:chic1}
Comparison of the integrands
$\frac{f_{ls}\mathcal{M}_{l's',ls}}{(z_X-E)}$ for $D^0\bar D^{0*}\to
\chi_{c1}\pi^0$~(solid) and $D^+\bar D^{-*}\to
\chi_{c1}\pi^0$~(dashed), when $z_{X(3872)}$ is chosen at $3.8716$GeV
as an example.}
\end{center}%
\end{figure}

Because the $X(3872)$ is very close to the $D^0\bar D^{0*}$ threshold,
the $1/(z_X-E)$ term will greatly enhance the contributions of $f_{l
s}\mathcal{M}_{l's',ls}$ near the $D^0\bar D^{0*}$ threshold, and it
also leads to extreme suppression of the contributions of the $D$-wave
$D\bar D^{*}$ components. As an example,
$\frac{f_{ls}\mathcal{M}_{l's',ls}}{(z_X-E)}$ for $S$-wave $D^0\bar
D^{0*}$ or $D^+\bar D^{-*}$ to $P$-wave $\chi_{c1}\pi^0$ is plotted in
Fig.~\ref{fig:chic1}. Since the flavor wave functions of $\pi^0$ is
$(\bar u u-\bar d d)/\sqrt{2}$, the cancellation naturally happens
between the neutral charmed states $D^0\bar D^{0*}$ and the charged
$D^+\bar D^{-*}$ components, which is similar to that of $X(3872)\to
J/\psi\rho$~\cite{Zhou:2017txt}. One could find that the contributions
of $D^0\bar D^{0*}$ and $D^+\bar D^{-*}$ in the large momentum region
will cancel each other and the contribution near the $D^0\bar D^{0*}$
threshold will be dominant.

In this calculation, the decay rates of $X(3872)$ to $\chi_{cJ}\pi^0$
for $J=0,1,2$ turn out to be very small, of the order of $10^{-7}$ GeV, with a
ratio $\Gamma_0:\Gamma_1:\Gamma_2=1.5:1.3:1.0$. This ratio is
comparable with the effective field theory calculations in
Refs.\cite{Dubynskiy:2007tj,Fleming:2008yn}. Our calculation also
suggests that the magnitude of the decay rates $\chi_{cJ}\pi^0$ might
not be large even if the $D^0\bar{D}^{0*}$ component is dominant. In
Refs.\cite{Dubynskiy:2007tj,Fleming:2008yn} a factor determined by the
internal dynamics cannot be determined, so they did not present the
magnitudes of such decay rates.

At the same time, we could also calculate the decay rates to
$J/\psi\pi^+\pi^-$ and $J/\psi\pi^+\pi^-\pi^0$ by assuming the final
states $\pi^+\pi^-$ and $\pi^+\pi^-\pi^0$ produced via $\rho$ and
$\omega$ resonances, respectively.  The interference of neutral and
charged $DD^*$ components in $X(3872)\to J/\psi\rho$ are destructive,
while it is constructive in $X(3872)\to J/\psi\omega$. For simplicity, we describe
the $\rho$ and $\omega$ resonances by their Breit-Wigner distribution
functions~\cite{Zhang:2009bv}, and then obtain
\bqa
\Gamma_{J/\psi\pi\pi}=\int_{2m_\pi}^{m_X-m_{J/\psi}}{ \sum_{l,s}}\frac{|F{_{l,s}}(X\rightarrow J/\psi\rho)|^2\Gamma_\rho}{(E-m_\rho)^2+\Gamma_\rho^2/4}\mathrm{d}E,\nonumber\\
\Gamma_{J/\psi\pi\pi\pi}=\int_{3m_\pi}^{m_X-m_{J/\psi}}{ \sum_{l,s}}\frac{|F{_{l,s}}(X\rightarrow J/\psi\omega)|^2\Gamma_\omega}{(E-m_\omega)^2+\Gamma_\omega^2/4}\mathrm{d}E,\nonumber\\
\label{eq:GammaX}
\eqa
in which the lower limits of the integration are chosen at the
experiment cutoffs as in Refs.~\cite{Abe:2005ix,delAmoSanchez:2010jr}.

The obtained decay width of $J/\psi\pi^+\pi^-$ is of the order of keV, and the ratio of decay rates to $X(3872)\to \chi_{c0}\pi^0$, $\chi_{c1}\pi^0$, $\chi_{c2}\pi^0$, $J/\psi\pi^+\pi^-$, and $J/\psi\pi^+\pi^-\pi^0$ is about $1.5:1.3:1.0:16:26$.

This calculation is based on the Barnes-Swanson model and the meson wave
functions are approximated by the simple harmonic oscillator wave functions for computing the
space overlap factor. This may introduce the ``prior-post"
discrepancies \cite{Barnes:1991em,Barnes:2000hu}
which are shown in the right graph in Fig.~\ref{fig:chic1}. Despite of
these discrepancies, 
the order of magnitudes of the prior and post contributions are
similar and we take the average of them as the final amplitudes.
Thus we would
expect that  the absolute magnitude of the decay width is just a rough
estimation and only provides an order of magnitude estimate. In
this calculation, the decay rate of $X(3872)$ to $\chi_{cJ}\pi^0$ is much
smaller than to $J/\psi\pi^+\pi^-$. We think the ratio is reasonable in
the mechanism proposed in this paper, because the final
$\chi_{cJ}\pi^0$ states could only appear in $P$ wave, while the
$J/\psi\rho$ states could appear in $S$ wave. Usually, the higher
partial waves will be suppressed. Furthermore, the phase space of
$\rho\to\pi^+\pi^-$ will enlarge the decay width of $X(3872)\to
J/\psi\pi^+\pi^-$. In \cite{Dong:2009yp},  in the
pure molecule picture, an effective field theory calculation gives
larger decay widths of $X(3872)$ to $\chi_{cJ}\pi$. However, their
branching fraction of ${\mathcal{B}(X(3872)\to\chi_{c1}\pi)}:{\mathcal{B}(X(3872)\to J/\psi\pi^+\pi^-)}$ is about ${(10.2\sim 16.4)}:{(45\sim 54)}$, which also implies a much smaller decay rate to $\chi_{c1}\pi$ than to $J/\psi\pi^+\pi^-$.
In our calculation, the $\chi_{c1}(2P)$ component in the $X(3872)$, which plays an important
role in the short range production processes, is expected to
contribute little in the long range decay processes and is ignored.
As a further check, by using the estimated value of the
partial decay width from pure $\chi_{c1}(2P)$ to $\chi_{c1}\pi^0$, which is about
$0.06$ keV~\cite{Dubynskiy:2007tj}, and considering the portion of $\chi_{c1}(2P)$ in $X(3872)$ to be
about $1/10$, its contribution to the decay width is about $6$~eV,
about two orders of magnitude smaller than the contribution from
$D\bar D^*$. Thus, this assumption is still valid.

In addition, the ratio  $\frac{\mathcal{B}(X(3872)\to
J/\psi\pi^+\pi^-\pi^0)}{\mathcal{B}(X(3872)\to J/\psi\pi^+\pi^-)}$ in
our calculation is about 1.6,
which is comparable with the measured result $1.0\pm 0.4\pm 0.3$
by Belle\cite{Abe:2005ix}, $0.8\pm 0.3$ by
\textsl{BABAR}~\cite{delAmoSanchez:2010jr} and $1.6^{+0.4}_{-0.3}\pm
0.2$ by BESIII~\cite{Ablikim:2019zio}. Thus, the isospin breaking
effect can be reproduced in this calculation as
in~\cite{Zhou:2017txt}.

In summary, by combining the extended Friedrichs scheme and the
Barnes-Swanson model, we make a calculation of the decay rates of
$X(3872)\to \chi_{c0}\pi^0$, $\chi_{c1}\pi^0$, $\chi_{c2}\pi^0$,
$J/\psi\pi^+\pi^-$, and $J/\psi\pi^+\pi^-\pi^0$ in a unified framework,
and find that the relative ratio will be about $1.5:1.3:1.0:16:26$. The decay
rate of $X(3872)$ to $\chi_{c1}\pi^0$ is one order of magnitude smaller than
$X(3872)$ to $J/\psi\pi^+\pi^-$ in this calculation. Our result is
smaller than the central value measured by
BESIII~\cite{Ablikim:2019soz}, but we noticed that the result of
BESIII has sizable uncertainties, and more data are needed to increase
the statistics and reduce the error bar. In Belle's
experiment, no significant evidence of $X(3872)$ signal was observed in $B^+\to\chi_{c1}\pi^0
K^+$~\cite{Bhardwaj:2019}, though its upper limit of $\frac{\mathcal{B}(X(3872)\to \chi_{c1}\pi^0)}{\mathcal{B}(X(3872)\to
J/\psi\pi^+\pi^-)}$ does not contradict with BESIII's result.
Recently, the Belle~II has started to accumulate data with higher statistics and it is expected that more accurate measurements could be obtained in the future.

\begin{acknowledgments}
Helpful discussions with Cheng-Ping Shen, Dian-Yong Chen, and Hai-Qing
Zhou are appreciated. This work is supported by China National Natural
Science Foundation under Contracts No. 11975075, No. 11575177, No. 11105138 
and by the Natural Science Foundation of Jiangsu Province
of China under Contract No. BK20171349. Z.X. also thanks S.Y. Zhou for
providing computational resources.
\end{acknowledgments}

\bibliographystyle{apsrev4-1}
\bibliography{Ref}

\end{document}